\documentclass[conference]{IEEEtran}
\IEEEoverridecommandlockouts
\usepackage{cite}
\usepackage{amsmath,amssymb,amsfonts}
\usepackage{algorithmic}
\usepackage{graphicx}
\usepackage{textcomp}
\usepackage{xcolor}
\usepackage[linesnumbered,boxed,lined,commentsnumbered, resetcount]{algorithm2e}
\usepackage{booktabs}
\usepackage{multirow}

\usepackage{subfigure}
\def\BibTeX{{\rm B\kern-.05em{\sc i\kern-.025em b}\kern-.08em
    T\kern-.1667em\lower.7ex\hbox{E}\kern-.125emX}}
\begin{document}

\title{Coverage Explorer: Coverage-guided Test Generation for Cyber Physical Systems \\ }

 \author{\IEEEauthorblockN{Sanaz Sheikhi}
 \IEEEauthorblockA{
 \textit{Stony Brook University}\\
 Stony Brook, New York, USA \\
 ssheikhi@cs.stonybrook.edu}
 \and
 \IEEEauthorblockN{Stanley Bak}
 \IEEEauthorblockA{
 \textit{Stony Brook University}\\
 Stony Brook, New York, USA \\
 sbak@cs.stonybrook.edu}
 }

\maketitle

\begin{abstract}
Given the safety-critical functions of autonomous cyber-physical systems (CPS) across diverse domains, testing these systems is essential. While conventional software and hardware testing methodologies offer partial insights, they frequently do not provide adequate coverage in a CPS. 
In this study, we introduce a testing framework designed to systematically formulate test cases, effectively exploring the state space of CPS. This framework introduces a coverage-centric sampling technique, coupled with a cluster-based methodology for training a surrogate model. 
The framework then uses model predictive control within the surrogate model to generates test cases tailored to CPS specifications. To evaluate the efficacy of the framework, we applied it on several benchmarks, spanning from a kinematic car to systems like an unmanned aircraft collision avoidance system (ACAS XU) and automatic transmission system. Comparative analyses were conducted against alternative test generation strategies, including randomized testing, as well as falsification using S-TaLiRo.

\end{abstract}


\section{Introduction}

The emergence of Autonomous Cyber Physical Systems (CPS) is making a tremendous change our daily lives. Applications such as self-driving cars, smart infrastructure, and industrial automation highlight the importance of ensuring the reliability and safety of these systems through testing and verification. 

In verification techniques, methods such as reachability analysis and theorem proving offer assurances superior to simulation based testing. However, their applicability is limited to relatively uncomplicated systems, and dealing with dimensionality and scalability inherent to CPSs is a significant challenge. Moreover, techniques like reachability analysis, while pivotal, need to deal with the subtle task of determining the precise set of reachable states, often ending up in approximations due to feasibility constraints. The accuracy of these approximations depends significantly on the  tuning of user-defined algorithm parameters, such as the time step size. Another variant of CPS verification, falsification, solves stochastic optimization problems to identify counterexamples. Nevertheless, this endeavor encounters scalability challenges, particularly concerning high-dimensional nonlinear CPSs.

Rigorous testing guarantees that autonomous CPSs operate as intended, thereby mitigating the risk of malfunctions that could lead to accidents or system failures. In practice, testing and simulation emerge as the predominant methods employed to ensure CPS correctness due to their ability to scale to complex systems.
Traditional software test methodologies are characterized by excessive costs, management challenges, and inefficacy in verifying CPS behavior. Moreover, their inability to handle the evolving CPS control requirements and dynamic plant parameters exacerbates their limitations.

This paper proposes a methodology for test case generation tailored to autonomous CPS requirements, focusing on state space coverage. This work introduces a sampling technique combining the rejection sampling method with the coverage notion. Furthermore, a data-driven model training approach for building models of black-box systems focusing on state space coverage is proposed. Finally, a coverage-guided test case generation utilizing model predictive control (MPC) in conjunction with the coverage-guided sampling method and coverage-guided model training is presented as the main contribution.

State sampling is pivotal to our search-based approach. Rather than randomly sample from a uniform distribution, our methodology intertwines the state space coverage concept with rejection sampling. This integration amplifies the likelihood of sampling from unexplored areas while diminishing the probability of replicating previously sampled states. In this context, rejection sampling targets areas where test cases have yet to be generated. %This is achieved by defining the main function as an integration of kernel functions superimposed on previously sampled states. Consequently, this coverage-based function guides the sampling of new states, ensuring their distance from these kernels using rejection sampling.

Our coverage-guided methodology applies to both white-box and black-box systems. When a white-box system is available, we can bypass the model training phase and proceed directly with coverage-guided test case generation. However, our focus in this study leans towards black-box CPSs, characterized by complex and unknown dynamics. To achieve this goal, we deploy the Koopman operator, a data-driven technique to build linear models from non-linear systems. To gather training data, rather than generating random data traces out of haphazard simulations, our methodology proposes a data generation technique that iteratively generates data traces expanding over the state space. At each iteration, a more precise model is developed, and the model is used to generate training data based on a coverage-guided mechanism for the next iteration. This incremental approach contrasts with the conventional method, where random simulations generate all training data in a singular endeavor. The difference is that randomly generated training data covers specific space segments, leaving vast areas unexplored and making the model biased to limited segments. In contrast, our training method is focused on learning the state space incrementally and covering more areas. Our methodology utilizes clustering to handle diverse data traces and deploys computational geometry to find training data bounds to be used in our sampling method.

Following model training, the test case generation phase begins. We sample target states and utilize MPC, operating on the trained model to minimize the distance to the target state and generate control inputs, steering the system toward these target states and contributing to state space coverage.

 To assess our work, we applied the proposed methodology across a spectrum of CPS, ranging from the simplicity of kinematic cars to the complexities embodied by automatic transmissions. We performed comparisons against a baseline uniform random test method and a falsification tool, S-TaLiRo.

In the following, section \ref{sec:related_work} provides a literature review. Section \ref{sec:preliminaries} provides the preliminaries, and section \ref{sec:Methodology} discusses the methodology. Section \ref{sec:evaluation} evaluates the proposed approach, and section \ref{sec:conclusion} concludes the paper.
\section{RELATED WORK}
\label{sec:related_work}
Numerous techniques have been proposed to verify the behavior of CPSs. Tools like NNV \cite{tran2020cav2}, OVERT \cite{SidraneOvert2021}, ReachNN* \cite{ChaoReachNN2019}, and VenMAS \cite{AkintundeVenMas2022} have been developed for verifying CPS equipped with Neural Network Control (NNCS). Despite these efforts, verifying neural networks remains a challenging task due to its NP-complete nature \cite{katz2017reluplex}. Notably, existing methods need help with large-scale and highly complex neural networks \cite{LiuASO2017} \cite{DvijothamDualApproach2018} \cite{RajeevVerification2011}. Moreover, current research has mainly focused on verifying neural networks' pre/post conditions, making it still challenging to assess their behavior within CPS contexts \cite{SouradeepRegressivePolynomial2019} \cite{ARCH_COMP20_Category_Report_Artificial}.

Falsification is a widely employed method for testing and verifying CPSs \cite{annpureddy2011s} \cite{donze2010breach}. Tools like Breach \cite{donze2010breach} and S-TaLiRo \cite{annpureddy2011s} \cite{FainekosP09} \cite{fainekos2009robustness} utilize stochastic optimization techniques to generate input test cases. These inputs are designed to violate specifications (e.g. safety specifications). The key distinction lies in the objectives: falsification aims to discover trajectories that breach safety specifications, whereas testing methods are geared towards exploring the system's state space, aiming to capture the entirety of the system's behavior. Falsification can get stuck in local minima, making it uncertain if the system is safe. In contrast, coverage-guided testing explores the entire state space, avoiding this issue. Combining both methods can mitigate their limitations and provide more reliable results.

Reachability analysis verifies CPSs behavior by computing reachable sets of states from initial states over a specific time \cite{althoff2008verification} \cite{althoff2010reachability} \cite{althoff14_online_journal}. This method is often combined with other techniques for falsifying hybrid and closed-loop control systems \cite{bogomolov2019falsification} \cite{goyal2020neuralexplorer}. However, the complex dynamics of CPS can complicate reachable set computations. Conservative over-approximation may help but compromises the feasibility of reachability analysis for CPS verification.

The star discrepancy metric ensures accurate representation of hybrid and dynamical systems by measuring the uniformity of point distribution within a set. It detects irregularities and gaps, highlighting clustered or sparse points for system analysis. Researchers have explored techniques leveraging star discrepancy's sensitivity to identify critical system areas, enhancing coverage-guided test generation efficiency \cite{Nahhal2007TestCoverage} \cite{Nahhal2007GuidedRandomizedSimulation} \cite{Dang2008SensitiveStateExploration} \cite{Barbot2020Falsification}. However, its computational cost, especially in high dimensions, hampers real-time applicability. While it offers uniformity insights, it lacks contextual system details, necessitating further analysis to link metric results to system behavior.
\section{Preliminaries}
\label{sec:preliminaries}
%Our proposed method utilizes well-known techniques like the K-mean algorithm for classification, the convexHull concept from computational geometry for state space bound identification, model predictive control (PMC) from control theory, and rejection sampling. 
This section will provide a brief introduction to CPS coverage score and Koopman operator linearization that are used in our methodology, as well as a problem statement.

\subsection{CPS COVERAGE SCORE}
\label{subsec:CPS_Score}
\begin{figure}[!b]
  \centering
  \includegraphics[width=0.70\columnwidth]{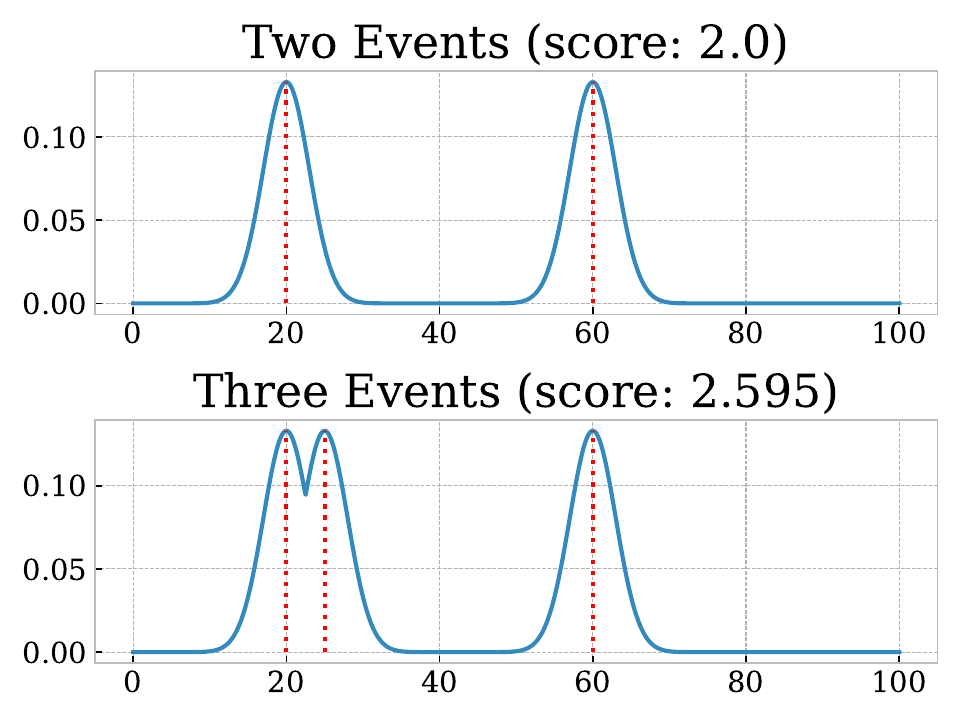}
  \caption{Our proposed CPS coverage metric increases as more 
           interesting events are generated, although events close 
           in the objective space do not increase the score as much 
           as separate events.}
  \label{fig:score}
\end{figure}

Understanding coverage is crucial to guide the state space search. Classic software coverage metrics like branch or line coverage fall short in CPSs. To address this issue, we employ a CPS coverage metric, aka CPS coverage score, from \cite{SheikhiCPSFuzz2022}, assessing state similarity. Three key properties characterize the CPS coverage metric: adding more states never decreases the metric, identical states do not increase the metric, and similar states have a lower impact than dissimilar states.
 
The states serve as input for computing the coverage score $\mathcal{S}$, which outputs a real-valued scalar. The function $\mathcal{S}$ operates on a finite set of sensed states, denoted as $\textnormal{Set}[Y]$, and maps it to real numbers ($\mathcal{S} : \textnormal{Set}[Y] \rightarrow \mathbb{R}$). 

To calculate the coverage score, the user supplies an Objective Space Projection Function denoted as $\mathcal{P} : Y \rightarrow \mathbb{R}^n$. This function maps the sensed state to a $n$-dimensional Euclidean space known as the \emph{objective space}, where coverage measurement occurs. Additionally, Objective Space Exploration Limits $\mathcal{B}\in\mathbb{R}^{2n}$ are specified, establishing boundary constraints within the objective space.

CPS coverage score is computed by mapping each state to the objective space using $\mathcal{P}$, applying a kernel function to measure state similarity, and integrating the maximum of the kernels within the objective space bounds $\mathcal{B}$. The kernel, a hyper-parameter function, assesses state similarity and is represented as an $n$-dimensional normal distribution $\mathcal{N}(\mu,\,\sigma^2)$, where $\mu$ denotes the point in the objective space corresponding to each state and $\sigma$ represents a constant hyper-parameter.
The CPS coverage score, is calculated by:
% $$\mathcal{S}(\textnormal{Set}[Y]) = \int_{B} ~~~~ \max_{~~y \in \textnormal{Set}[Y]} \mathcal{N}(\mathcal{P}(y),\,\sigma^2)(b)  ~~ \mathrm{d}b$$
% \label{eq:coverage}

\begin{equation}
\mathcal{S}(\textnormal{Set}[Y]) = \int_{B} \max_{y \in \textnormal{Set}[Y]} \mathcal{N}(\mathcal{P}(y), \sigma^2)(b) \, \mathrm{d}b
\label{eq:cps_coverage}
\end{equation}

Figure~\ref{fig:score} illustrates a straightforward example of CPS coverage score computation. The objective space is a single dimension along the $x$-axis, bounded by $\mathcal{B} = (0, 100)$. At the top, two states (indicated by red dotted lines) are located at 20 and 60 in the objective space. Due to the significant distance concerning the $\sigma$ standard deviation hyper-parameter, the score is approximately two. At the bottom, a third state mapped to 25 in the objective space is added, raising the score to about $2.595$. Because this state was close to the other state at 20, the score does not reach 3, as it would if the count of  states was solely considered. The figure calculates each case's score (CPS coverage metric) by integrating the blue curve.

\subsection{Koopman Operator Linearization}
\label{subsec:koopman}
Many CPSs are complex nonlinear systems, posing challenges in analysis and verification. Unlike linear systems, nonlinear ones lack straightforward methods for analysis. We opt for a linear approximation of these nonlinear systems to tackle this challenge. The concept involves fitting the system's nonlinear dynamics into a higher-dimensional linear space. By employing a higher-dimensional linear system, the Koopman operator approximates the dynamic behavior of the nonlinear system  \cite{KoopmanHmiltonian1931}.

This approach is advantageous in our testing process because it allows us to optimize the model using MPC, benefiting from efficient optimization algorithms designed for linear models. 

Although an infinite-dimensional linear system can model nonlinear behavior using observables, practical limitations requires us working with a finite set of observables. To ensure meaningful results, accurate approximation is crucial during Koopman operator linearization, achieved through observable functions. Observable function selection can involve Fourier features or neural networks \cite{Han2020DeepLearningKoopman} \cite{Rahimi2007RandomFeatures} \cite{Yeung2019DeepNeuralKoopman}. Our approach opted for random Fourier features as observable functions, offering a systematic and high-accuracy process. Unlike ad hoc finite-dimensional spaces, random Fourier features utilize powerful kernel techniques from machine learning \cite{Kim2005kernelInduced} \cite{Scholkopf2001LearningWithKernels}, constructing a computationally manageable mapping across an infinite-dimensional feature space. We nonlinear systems represented as follows:
\begin{multline}
x_{k+1} = f(x_{k}, u_{k}) \quad with \quad  x_{k} \in \mathbb{R}^n, \\
u_{k} \in \mathbb{R}^w, \quad f: \mathbb{R}^n \times \mathbb{R}^w \longrightarrow \mathbb{R}^n
\label{eq1}
\end{multline}

For Koopman linearization observable functions $g_{i}: R^n \longrightarrow R$ are required where the dynamics of the resulting variable $g_{i}$ is linear:

\begin{equation} \label{eq2}
g(x_{k+1}) = Ag(x_{k}) + Bu_{k} \quad with \quad  A \in R^{m \times m}, \quad B \in R^{m \times w},
\end{equation}

In this context, the collective observable function is denoted as  $g(x_{k}) = [g_{1}(x_{k}) ... g_{m}(x_{k})]^T$. The variables $g_{i}(x_{k})$ are derived from the original system state \(x_{k}\), and thus, the linear system described by (\ref{eq2}) captures the dynamic patterns of the original system outlined in (\ref{eq1}). Typically, the count of observables $m$ significantly exceeds the dimensionality of the original system $n$.

The most accurate representation of the system, matrices $A$ and $B$ can be obtained through extended dynamic mode decomposition as outlined in \cite{Williams2015ADA}, utilizing traces of the original system. These traces can originate from black-box simulations or real system behavior measurements, eliminating the necessity for a specific model (\ref{eq1}) of the original system. This characteristic is pivotal, rendering the Koopman operator invaluable for data-driven methodologies.

\subsection{Problem Statement}

Testing CPSs is crucial due to their complex continuous state space and time-varying inputs. As a motivating example, consider a 3D point mass system where the state includes position. Eq. \ref{eq:toy_dynamic} defines the dynamics.

\begin{equation}
\begin{aligned}
\bar{x}  = \bar{x} + v_x, \\
\bar{y} = \bar{y} + v_y, \\
\bar{z} = \bar{z} + v_z
\label{eq:toy_dynamic}
\end{aligned}
\end{equation}

\begin{figure}[t]
    \centering
    \subfigure[Random test cases]{\includegraphics[width=0.49\columnwidth]{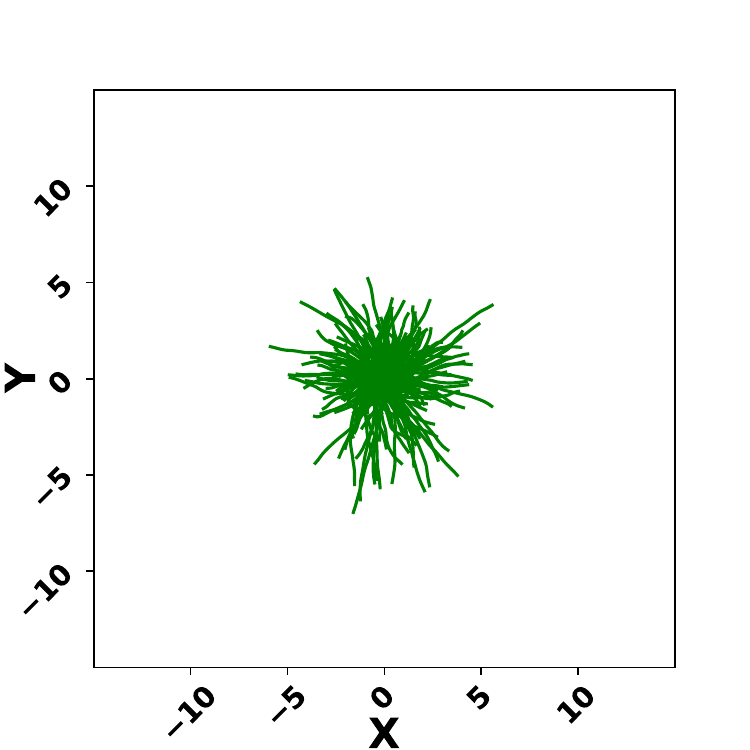}}
    % \hfill
    \subfigure[Coverage-guided test cases ]{\includegraphics[width=0.49\columnwidth]{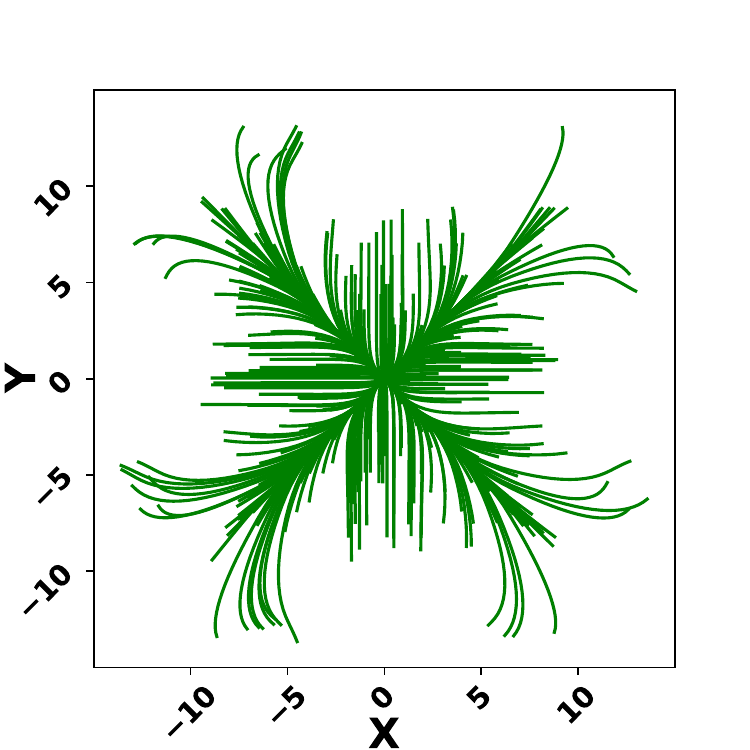}}
    \caption{State space coverage}
    \label{fig:toy_example}
\end{figure}

The intuitive goal of our approach is to select test cases (inputs) that guide the point mass to a variety of different places in the $3D$ state space, originating from $(0,0,0)$. 
As a preview of our results, we show the performance of a random input baseline compared with the framework we will present. Figure \ref{fig:toy_example} compares state trajectories (projected to $2D$ space for visualization), revealing that our method better explores the set of possible states.
We next formalize this intuition on a general CPS.

\textbf{Problem Statement}: The state vector of the system is $x \in \mathbb{R}^n$, where $n$ is the number of variables in the system. 
The input is $u \in \mathbb{R}^w$, where $w$ is the number of inputs.
An input vector is a sequence of inputs at each time step, $\bar{u} = u_{0}, u_{1}, \ldots u_N$.
%
%Let $T \in \mathbb{Z^+}$ be the number of steps being analyzed. 
%
The system executes in discrete time with $x_{k+1} = f(x_k, u_k)$.
Given initial states $x_{0,1}, x_{0,2}$,\ldots $\in \mathbb{R}^n$, and a time bound $T \in \mathbb{Z}^{\geq 0}$, 
the goal is to create input vectors $\bar{u}_1, \bar{u}_2$, \ldots that maximize coverage, as defined by the CPS coverage score from Equation~\ref{eq:cps_coverage}.
%%%%%%%%%%%%%%%%%%%%%%%%%%%%%%%%%%%%%%%
\section{METHODOLOGY}
\label{sec:Methodology}

\begin{figure} %[htbp]
  \centering
  \includegraphics[width=\columnwidth]{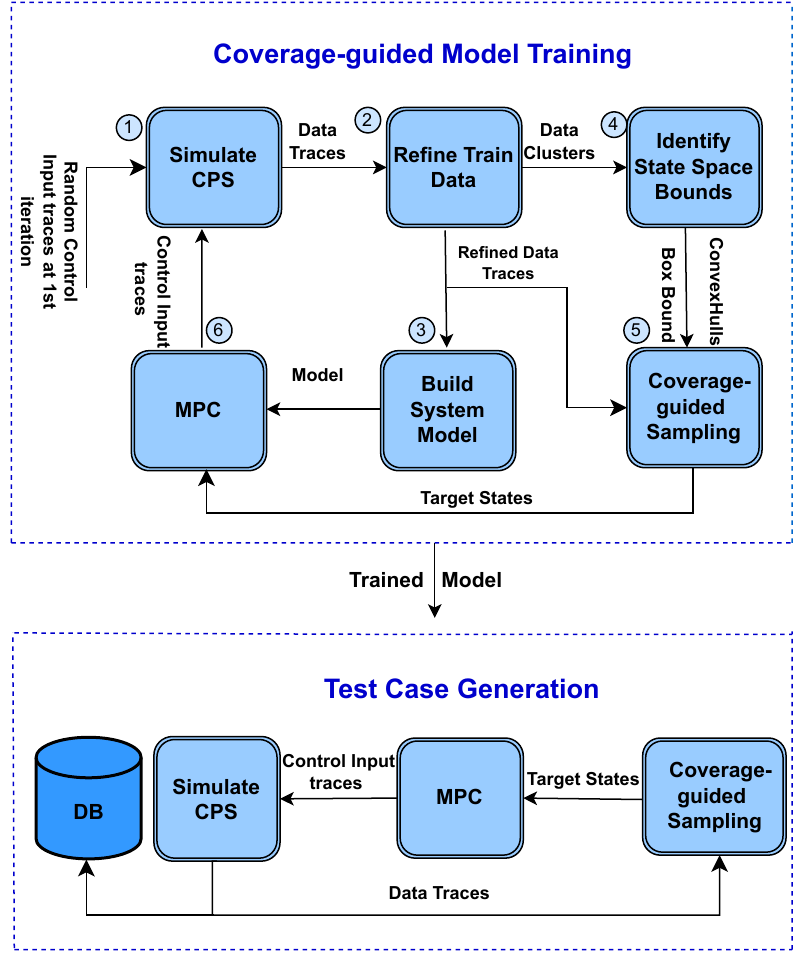}
  \caption{Coverage Explorer Overview}
  \label{fig:overview}
\end{figure}

To achieve the goal of coverage-guided test generation, we use a two-phase methodology as shown in Fig. \ref{fig:overview}, which consists of Model Training and Test Case Generation.

\textbf{Model Training} proceeds as follows:

  \begin{enumerate}
      \item \textbf{Generating Random  Data Traces}: The process starts with system simulations using random control inputs, generating state trajectories as output, which, along with the corresponding control input sequences, are referred to as data traces.

     \item \textbf{Refining Training Data}: In each model training iteration, we analyze existing training data traces from previous iterations and new simulations. State trajectories of these data traces are clustered, allowing selection of  a specific number of the trajectories from each cluster to represent the state space effectively and remove similar trajectories.

     \item \textbf{Build Model}: In each iteration, we build a new model from the refined training data using the Koopman operator. We employ Auto-Koopman~\cite{lew2023autokoopman}, an automated tool that selects and tunes the Koopman process hyperparameters to maximize model accuracy.

     \item \textbf{Identifying State Space Boundaries}: We use convex hulls from computational geometry to determine a tight bound around the training data, which we use in our state sampling technique.

     \item \textbf{Coverage-Guided Sampling}: Sampling approach, which combines CPS coverage score (explained in \ref{subsec:CPS_Score}) with rejection sampling (aka acceptance-rejection sampling) to enhance state space coverage. This technique is used during both model training and test case generation.

     \item  \textbf{Generate Coverage-guided Data Traces}: We are employing MPC to generate coverage-guided data traces. In each model training iteration, MPC operates on the system model, optimizing a cost function based on target states sampled via the coverage-guided approach, steering the simulations toward low coverage areas.
  \end{enumerate}

%\vspace{5}

\textbf{Test Case Generation}: Test case generation starts after the last model training iteration. To generate test cases, we run simulations where the control inputs are issued by  MPC operating on the trained model. MPC cost function is defined to minimize the distance between the initial state and the sampled state (target). Following each simulation, output state trajectories are used to update the data structures of the coverage-guided sampling method to prevent the sampling of similar states. Also, these state trajectories are used for coverage measurement purposes.

\subsection{Training Data Refinement}

Black-box systems produce data that we use to model the system's behavior. Considering the system's complexity, good coverage of the 
 multi-dimensional state space is essential, and models must capture this complexity for accuracy. Inadequate coverage biases models, affecting predictions and hindering 
 generalization. Comprehensive coverage represents most states and scenarios, ensuring reliable predictions
 
 We aim to pick diverse data traces representative of the state space and expand the state space exploration at each model training iteration. To achieve this goal, we use K-means clustering and similarity analysis to refine data traces. Data-driven approaches, such as K-means clustering and dissimilarity selection, are instrumental—clustering groups similar data, managing the diversity, and enabling targeted analysis. By selecting dissimilar traces, redundant or highly similar data that add little value are pruned away, focusing computational cycles on representative data traces.
 
In practice, we initiate with a fixed number of clusters (K). At each iteration with the growing number of data traces, we increment the number of clusters slightly. 
This expansion involves clustering a combination of data traces from the previous iteration and new data from coverage-guided simulations.

Algorithm \ref{algo:Data_Refining} outlines the data refining process. It takes training data traces, cluster count, and selection rate as inputs. The algorithm initializes an empty list for selected data traces in line 1 and clusters the input data traces in line 2. From lines 3 to 13, it iterates over the clusters, calculates pairwise distances between traces (state trajectories), sorts them based on distance, and selects a specific number of dissimilar traces according to the selection rate. The selected traces are returned for model training and sampling method, along with the identified clusters for bound identification.

\SetAlgoNlRelativeSize{0}
\RestyleAlgo{ruled}
\SetKwComment{Comment}{/* }{ */}
\begin{algorithm}[]
\caption{Training Data Refinement}
\label{algo:Data_Refining}
\KwData{$K$ /*number of clusters*/}
\KwData{$input\_data$ /*training data traces*/}
\KwData{$rate$ /*data selection rate*/}
 $selected\_traces = \gets []$\;
 $clusters \gets Kmean(input\_data, K)$\;

\For{$cluster \in clusters$}{
 $data \gets get\_data\_traces(cluster)$\;
 $distances \gets []$\;
 $count \gets size(data)$\;
\For{$i \in range(count)$}{
\For{$j \in range(i + 1, count)$}{
$distance \gets get\_distance(data[i], data[j])$
$dist\_matrix \gets dist\_matrix \; U \; distance$
}
}
$curr\_traces \gets get\_dissimilar\_traces(data, dist\_matrix, rate)$
$selected\_traces \gets selected\_traces \; U \; curr\_traces $
}
$\texttt{return selected\_traces, clusters}$\;

\end{algorithm}
\RestyleAlgo{ruled}
\SetKwComment{Comment}{/* }{ */}

\subsection{Model Training}

Black-box systems present a significant challenge because of their complex nature and the need for more transparency into their internal structure. However, investigating these systems is vital for decision-making, optimization, and control. Traditional analytical methods often fall short, while data-driven techniques unveil patterns and behaviors. The data-driven approaches have become powerful tools, harnessing data to build accurate and reliable models for black-box systems.

We use the Koopman operator approximation approach, as explained in subsection \ref{subsec:koopman}, to create a surrogate model of the system using the refined training data traces. We employ AutoKoopman, an advanced system identification tool that automatically optimizes all hyper-parameters, ensuring accurate system models with globally linearized representations~\cite{lew2023autokoopman}. 

Algorithm \ref{alg:model_training} outlines the model training process. It takes inputs such as the number of training iterations, the number of simulations per iteration, and the initial number of clusters. In the first iteration, data traces are randomly generated for training (line 2), where the simulation type is random, so it does not require the system model. During each subsequent iteration, the training data is a combination of selected traces from the previous iteration and newly generated ones (line 4). These traces are refined using Algorithm \ref{algo:Data_Refining} (line 5). A new surrogate model is then constructed using the Koopman operator(line 6). Simulations run with MPC, which utilizes this model to generate new data traces (line 7). The iteration count is adjusted (line 8), and the number of clusters is increased (line 9) as more training data is acquired in subsequent iterations to maintain the representativeness of the data traces. Following the final iteration, the fully trained model is provided for use in the test case generation phase (line 11).

\RestyleAlgo{ruled}
\SetKwComment{Comment}{/* }{ */}
\begin{algorithm}[]
\caption{Model Training}
\label{alg:model_training}
\KwData{$iterations$ /*training iterations*/}
\KwData{$sim\_count$ /*simulations count*/}
\KwData{$cluster\_count$ /*clusters count*/}

 $new\_data\_traces \gets []$\;
 $data\_traces \gets simulate(random,  None, sim\_count)$\; 

  \While{$iterations \neq 0$}{
  $data\_traces \gets data\_traces\; U \; new\_data\_traces$\;
  $data\_traces \gets refine\_training\_data(data\_traces)$\;
  $model \gets auto\_koopman(data\_traces) $\;
  $new\_data\_traces \gets simulate(MPC, model, sim\_count)$\;
  $iterations \gets iterations - 1$\;
  $ cluster\_count \gets cluster\_count + 1$\;
 }
$\texttt{return model}$\;

\end{algorithm}
\RestyleAlgo{ruled}
\SetKwComment{Comment}{/* }{ */}

\subsection{Identifying state space boundaries}

The model training process relies on coverage-guided simulations using target states that are sampled to enhance state space coverage. Randomly sampling targets from anywhere in the state space could result in states too similar to previous simulations outputs, not contributing to coverage, or states far from the existing training data, leading to inaccurate predictions by the model. Hence, the challenge is to sample target states that are neither close nor distant from the previous data, providing a balance for effective coverage expansion.

To identify boundaries, we compute the convex hull of the data set. After selecting the set of training data traces at each iteration, we define a maximum and minimum box bound over the training data. Then, a tighter bound around all training data is established using a convex hull. The space between these two bounds is used to sample target states. This space achieves a compromise: the model has partial knowledge, preventing irrelevant outputs while allowing room for capturing new states and increasing training data coverage.

Observations reveal that enclosing all training data within a single convex hull leaves significant space inside the convex hull inaccessible, resulting in the loss of valuable data coverage. To tackle this problem, we obtain tighter bounds by placing each data cluster within a separate convex hull. Algorithm \ref{alg:get_boundaries} expresses the process of finding boundaries of the state space for sampling. The algorithm processes clusters of training data traces as input. It iterates over data clusters, computes the convex hull for each cluster (lines 3 to 8), and determines the box bound using the minimum and maximum values of each dimension (line 9).

\RestyleAlgo{ruled}
\SetKwComment{Comment}{/* }{ */}
\begin{algorithm}[h!]
\caption{Identifying training data bounds}
\label{alg:get_boundaries}
\KwData{$clusters:$ training data clusters}

$convex\_hulls \gets []$\;
$all\_training\_data \gets []$\;
\For{$cluster \in clusters$}{
$data \gets get\_culster\_data(cluster)$\;
$all\_training\_data \gets all\_training\_data \; U \; data $\;
$convexhull \gets compute\_convex\_hull(data)$\;
$convex\_hulls \gets convex\_hulls \; U \; convexhull$\;
}
$box\_bound \gets get\_box\_bound(all\_training\_data)$\;
$\texttt{return box\_bound, convex\_hulls}$\;

\end{algorithm}
\RestyleAlgo{ruled}
\SetKwComment{Comment}{/* }{ */}

\subsection {Coverage-guided Sampling}

Our coverage-guided testing method explores state space systematically by sampling from low-coverage areas, using a combination of  CPS coverage score and rejection sampling technique. 

The kernel function $\mathcal{S}(\textnormal{Set}[Y])$, discussed in section (\ref{subsec:CPS_Score}) measures sensed states similarity. By sensed states, we mean states generated by coverage-guided simulations. Our objective is to sample from $1 - \mathcal{S}(\textnormal{Set}[Y])$ to exercise uncovered states. On the other hand, this kernel function is a complex distribution to sample from.

So, we leverage rejection sampling to sample from this distribution because: 1) rejection sampling generates numbers within a probability distribution; if a number falls outside the bounds, it's discarded and repeated until a suitable number is obtained; 2) rejection sampling simplifies sampling from complex distributions by leveraging simpler, known distributions, a fundamental technique in Monte Carlo methods.

\begin{figure}[]
  \centering
  \includegraphics[width=0.7\columnwidth]{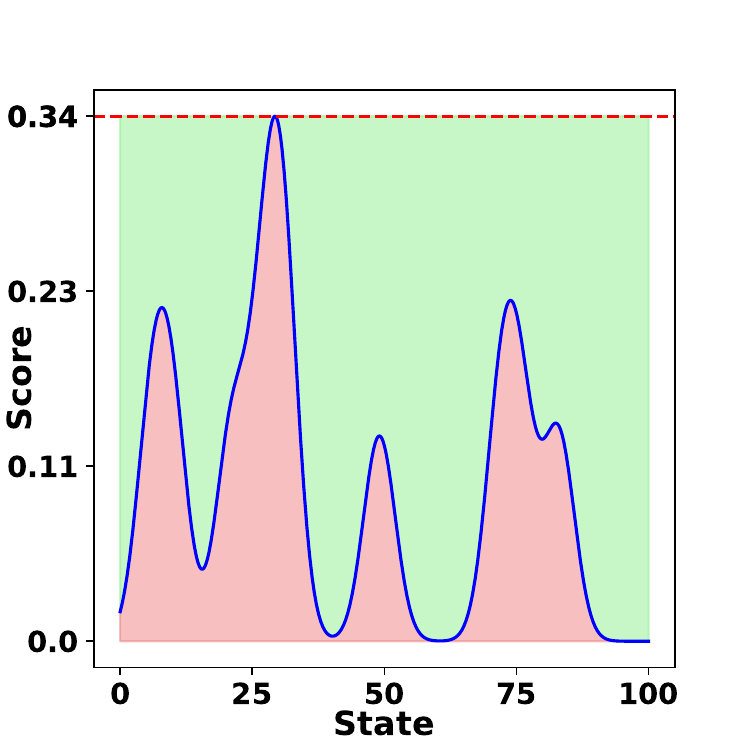}
  \caption{Coverage-guided sampling}
  \label{fig:coverage_sampling}
\end{figure}

Fig. \ref{fig:coverage_sampling} illustrates our rejection sampling technique. In this one-dimensional state example, the blue plot represents the kernel function defined over sensed states (the red area), and our goal is to sample from the green area (uncovered states). 

\section{EVALUATION}
\label{sec:evaluation}
    
We tested our approach across various CPS benchmarks, ranging from simple to complex systems such as the kinematic car, ACAS Xu, and automatic transmission. We compared our methodology's results against those of other test case generation techniques. This section provides an overview of benchmarks and an analysis of our experiments, including coverage-guided model training, sampling method, and test case generation technique. %Furthermore, we assessed test case generation for all benchmarks, quantifying state space coverage through CPS coverage scores, as elucidated in section \ref{subsec:CPS_Score}. We compared these coverage results with those obtained from test cases generated by a baseline approach, which generates uniformly random test cases, and STALIRO, a falsification tool.

\subsection{BENCHMARKS}
\label{subsec:Benchmarks}
% \textcolor{red}{Do we need description of the benchmarks? or just naming them with reference is enough?}
\begin{itemize}
    \item \textbf{Kinematic car}: A commonly used benchmark in the CPS domain is the kinematic single-track model of a car, sourced from \cite{Bastian2017Convex}. The simplified system dynamics for this autonomous car benchmark is outlined in Eq.(\ref{eq:kinematic}).
    \begin{equation}\label{eq:kinematic}
        \begin{split}
            & \dot{x_{1}} = u_{1} + w_{1}, \\
            & \dot{x_{2}} = u_{2} + w_{2}, \\
            & \dot{x_{3}} =x_{1}.\cos{x_{2}}, \\
            & \dot{x_{4}} =x_{1}.\sin{x_{2}}
        \end{split}
    \end{equation}
    
    The system states include velocity \(x_1 = v\), orientation \(x_2 = \phi\), and position \(x_3 = x, x_4 = y\) of the car. Inputs consist of acceleration \(u_1\) and normalized steering angle \(u_2\). Input constraints are \(u_1 \in [-9.81, 9.81] \, \text{m/s}^2\) and \(u_2 \in [-0.4, 0.4] \, \text{rad/s}\). Our experiments have no disturbances (\(w_1 = 0.0\) and \(w_2 = 0.0\)). The initial set is \(x_1(0) = 15.0\), \(x_2(0) = 0.0\), \(x_3(0) = 0.0\), and \(x_4(0) = 0.0\).

    \item \textbf{ACAS Xu}: The neural network compression of ACAS Xu, an air-to-air collision avoidance system for unmanned aircraft, described in \cite{lopez2022jat} \cite{Sheikhif2023arch} is a widely used benchmark. The system includes two aircraft, ownship and intruder, and takes inputs from both aircraft's states, denoted as \(I = \{\rho, \theta, \psi, v_{own}, v_{int}, \tau, a_{prev}\}\): %$\rho$ is the distance between ownship and intruder aircraft, $\theta$ is the angle to intruder w.r.t ownship heading, $\psi$ is the heading of intruder w.r.t ownship, $v_{own}$ is the ownship velocity, $v_{int}$ is the intruder velocity, $\tau$ is the time until loss of vertical separation and  $a_{prev}$ is the previous advisory.

    \begin{itemize}
        \item $\rho$: distance between ownship and intruder aircraft
        \item $\theta$: angle to intruder w.r.t ownship heading
        \item $\psi$: heading of intruder w.r.t ownship 
        \item $v_{own}$: ownship velocity 
        \item $v_{int}$: intruder velocity
        \item $\tau$: time until loss of vertical separation
        \item $a_{prev}$: previous advisory
    \end{itemize}
    
   The system's output comprises advisory turn commands for the ownship: % $SL$ as strong left turn at 3.0 deg/s, $WL$ as weak left  turn at 1.5 deg/s, $COC$ as clear of conflict (do nothing), $WR$ as weak right turn at 1.5 deg/s, $SR$ as strong right turn at 3.0 deg/s.
   
    \begin{itemize}
        \item $SL$: strong left turn at 3.0 deg/s
        \item $WL$: weak left  turn at 1.5 deg/s
        \item $COC$: clear of conflict (do nothing)
        \item $WR$: weak right turn at 1.5 deg/s 
        \item $SR$: strong right turn at 3.0 deg/s
    \end{itemize}

    This benchmark has two variants including Dubins and F-16 dynamics. In our study, we employed the benchmark with Dubins dynamics. Modeling aircraft with positions, velocities, and headings leads to a nonlinear model for each aircraft, represented in Eq. (\ref{eq:dubins_1}).

    \begin{equation}\label{eq:dubins_1}
    \begin{split}
        & \dot{x_{1}} = \dot{x} = v\cos({\omega}), \\
        & \dot{x_{2}} = \dot{y} = v\sin({\omega}), \\
        & \dot{x_{3}} = \dot{\omega} = u
    \end{split}
    \end{equation}

    the state of the aircraft needs to be converted into the input space of the neural network. The ACAS Xu neural networks, takes in the distance ($\rho$), angle to intruder w.r.t ownship heading ($\theta$), and heading of the intruder w.r.t. ownship ($\psi$). The conversion between the Dubins model parameters and the NNCS inputs can be performed using Eq. (~\ref{eq:dubins_NN}). Finally, before running the neural network, each input is scaled by  normalizing and offsetting it to a range of [-1, 1].

    \begin{equation}\label{eq:dubins_NN}
        \begin{split}
            & \rho =\sqrt{(x_{int} - x_{own})^2 + (y_{int} - y_{own})^2}, \\
            & \theta = \arctan{(\frac{y_{int} - y_{own}}{x_{int} - x_{own} + \rho})}- \psi_{own}\\
            & \psi = \psi_{int} - \psi_{own}
        \end{split}
    \end{equation}
    
    \item \textbf{Automatic transmission}: Automatic transmission model \cite{ARCH15:Benchmarks_for_Temporal_Logic} includes a controller responsible for choosing gears 1 to 4. This system exhibits both continuous and discrete behavior and has two inputs: throttle and brake. The brake input models variable engine load, like going uphill or downhill. The outputs are two continuous-time state variables: engine speed ($\omega$ in RPM) and vehicle speed ($V$ in mph). Initially, the vehicle is at rest ($v = 0$, $\omega = 0$). Output trajectories depend on $0 \leq throttle \leq 100$ and $0 \leq brake \leq 325$ inputs. The system is deterministic, producing the same output $y$ for the same input $u$. This is a modified version of the Automatic Transmission model from Mathworks, presented as a Simulink demo. It consists of $69$ blocks, $2$ integrators, $3$ look-up tables, $3$ $2D$ look-up tables, and a Stateflow chart.
\end{itemize}

\subsection{Evaluation of the coverage-guided model training}

    \begin{figure*}[]
        \centering
        \subfigure[1st iteration]{\includegraphics[width=0.25\textwidth]{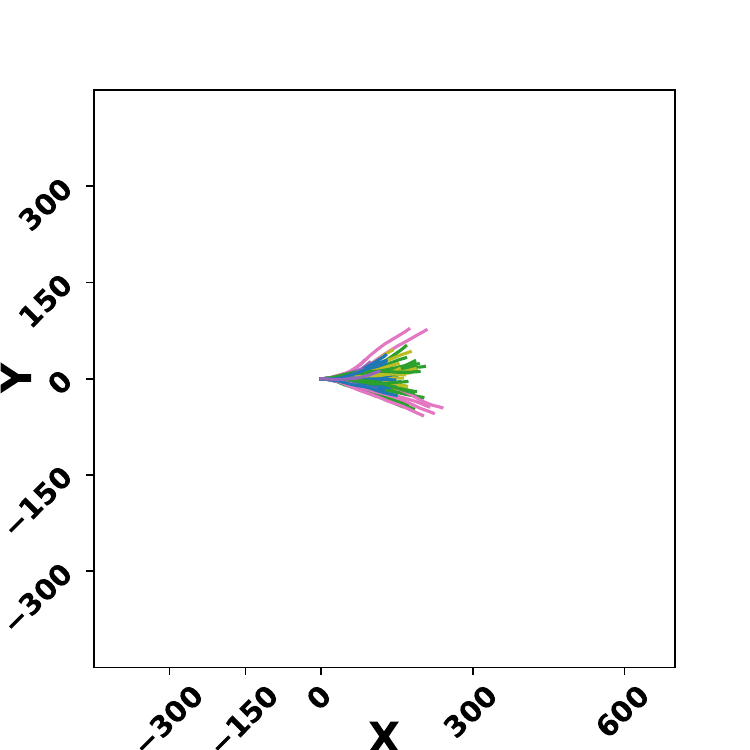}}\hfill
        \subfigure[2nd iteration]{\includegraphics[width=0.25\textwidth]{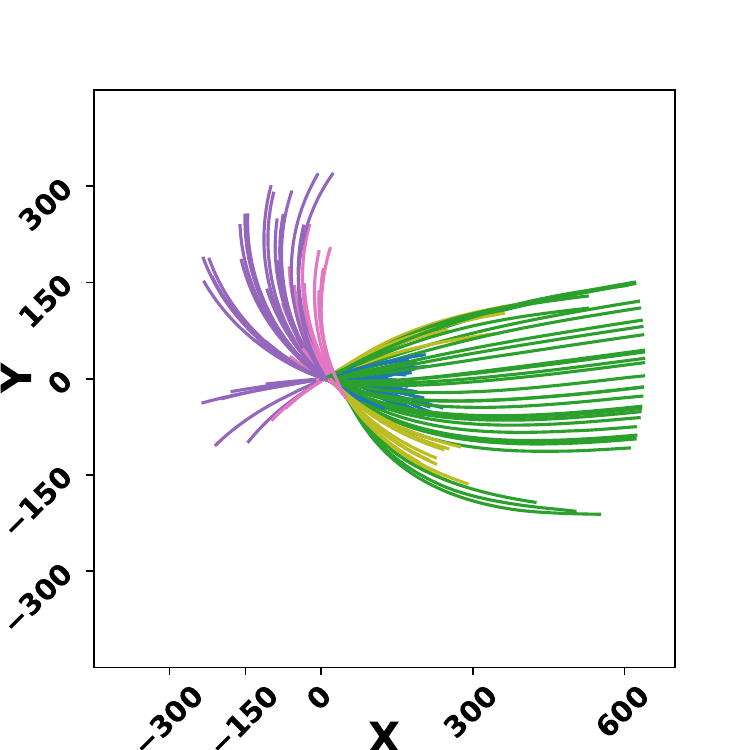}}\hfill
        \subfigure[3rd iteration]{\includegraphics[width=0.25\textwidth]{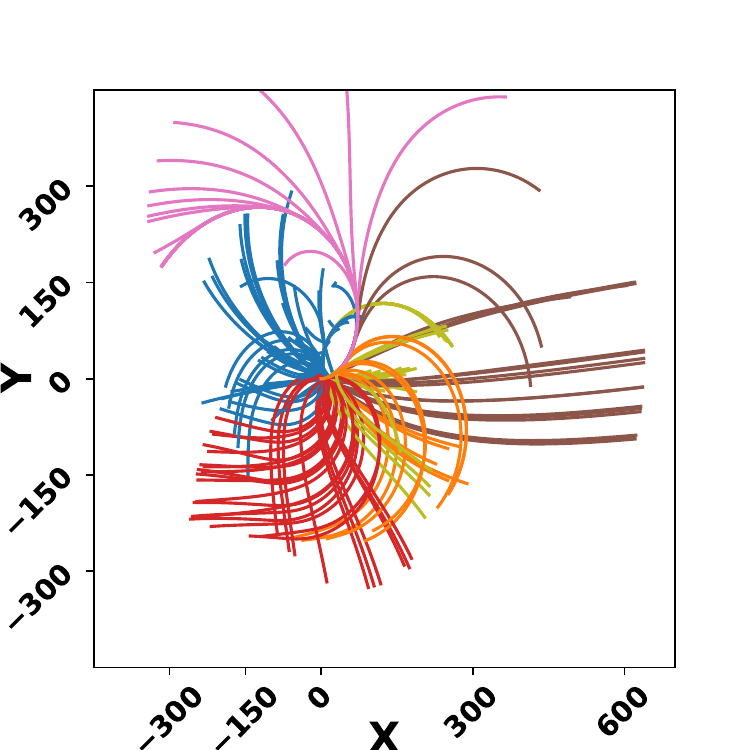}}\hfill
        \subfigure[4th iteration]{\includegraphics[width=0.25\textwidth]{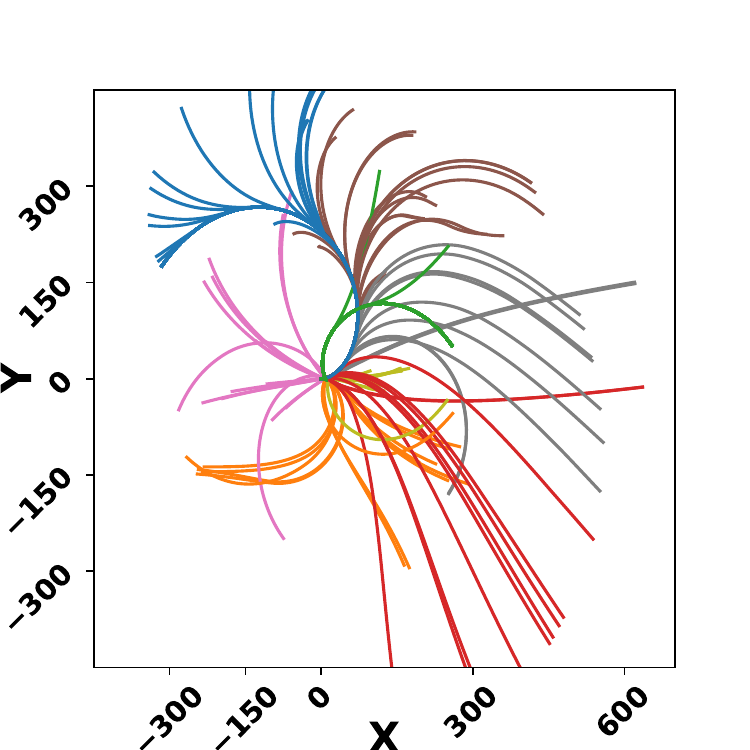}}\hfill
        % \subfigure[Caption for Figure 5]{\includegraphics[width=0.2\textwidth]
        % {Fig/cluster_6.png}}
        \caption{Visual representation of enhanced training data for kinematic car benchmark, expanding coverage across the state space through model training iterations. Additionally, illustrating clustered similar state trajectories aiding in selecting representative data and helping to establish precise bounds for target sampling.}
        \label{fig:clustering}
    \end{figure*}

\begin{table}
\centering
\renewcommand{\arraystretch}{1.5} % Increase row spacing
\begin{tabular}{cccccc}
\toprule
  \textbf{Iteration \#} & \textbf{Kinematic car} & \textbf{ACAS Xu} & \textbf{Auto Trans} \\
\midrule
\textbf{1}   & 272.42 & 2338.73 & 206.59 \\
\textbf{2}   & 1430.78 & 3958.95 & 346.82 \\
\textbf{3}  & 1628.22 & 4078.86 & 490.11 \\
\textbf{4}   & 2241.25 & 4531.86 & 536.56 \\
\bottomrule
\end{tabular}
\caption{Coverage improvement across model training iterations.}
\label{tab:trainingData_cov}
\end{table}

We conducted several experiments to assess our coverage-guided model training method. For all benchmarks, we trained the system model through several iterations using data traces generated through the coverage-guided simulations. 

Fig. \ref{fig:clustering} illustrates the evolution of kinematic car training data's scope across iterations, enhancing state space coverage for a more accurate model. Fig. \ref{fig:clustering}.a shows training data from simulations with random control inputs, showing a limited state space coverage. Observations revealed that continuously generating random data will not significantly alter this situation due to data trace similarity. So, applying a coverage-guided mechanism, we dramatically broaden the state space coverage at each iteration by adding data traces obtained from coverage-guided simulations and going through a refinement process. 

Fig. \ref{fig:clustering}.a to Fig. \ref{fig:clustering}.e visualize the coverage-guided model training process. CPS coverage score was computed for the training data at each iteration, indicating improved training on the state space. Table \ref{tab:trainingData_cov} shows the increasing state space coverage across iterations for all benchmarks.

\subsection{Evaluation of training data boundary identification technique}

\begin{figure}[t]
    \centering
    \subfigure[Bound identification with single convexhull]{\includegraphics[width=0.49\columnwidth]{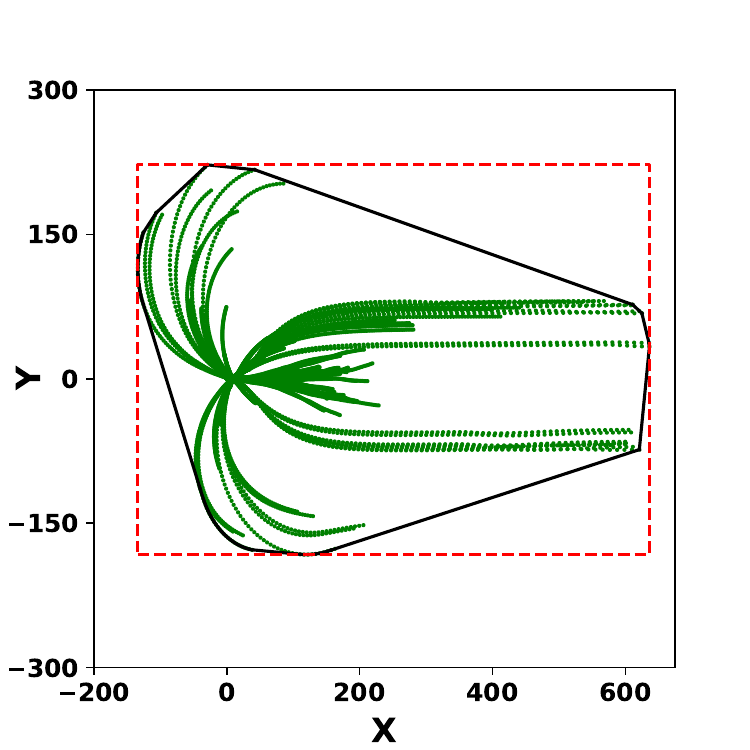}}
    \subfigure[Bound identification with multi convexhulls]{\includegraphics[width=0.49\columnwidth]{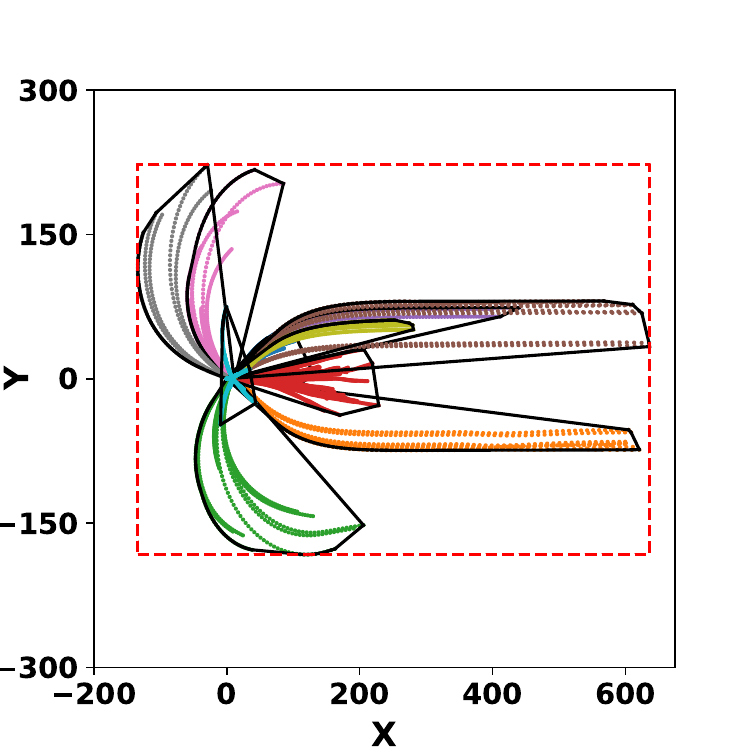}}
    \caption{Training data bound identification}
    \label{fig:convex_hulls}
\end{figure}

To evaluate the impact of training data boundary identification on sampling quality and coverage, we conducted experiments using three methods: a fixed box bound, a single convex hull, and multiple convex hulls. The fixed box bound, determined by the minimum and maximum of the training data, was offered maximum freedom for sampling. Fig.\ref{fig:convex_hulls} illustrates the space available for sampling between the red dashed box and the black solid lines (convex hull boundaries). Using multiple convex hulls provides a tighter bound around the training data and increases available space significantly compared to using single convex hull, which offers a looser bound. On the other hand, despite a wide sampling area within the fixed box bound, it does not enhance state space coverage. The reason is in sampling from the fixed box bound we do not care about state similarity and according to CPS coverage metric similar states similar states, do not contribute to the coverage score considerably.

In contrast, our convex hull-based technique allows sampling from uncovered areas, introducing new states. Table \ref{tab:boxbound_cov} displays CPS coverage scores for all benchmarks using the three methods. Employing multiple convex hulls considerably improves coverage-guided sampling, guiding the process toward unexplored state space regions; even a looser bound around training data outperforms a fixed box bound.

\begin{table}
\centering
\renewcommand{\arraystretch}{1.5} % Increase row spacing
\begin{tabular}{lccc}
\toprule
    & \textbf{Kinematic car} & \textbf{ACAS Xu} & \textbf{Auto Trans} \\ \midrule
\textbf{Fix Bound} & 1266.96 & 4252.97 & 388.92 \\
\textbf{Single Convex} & 1991.10 & 5614.93 & 402.83 \\
\textbf{Multi Convex} & 2176.58 & 6504.73 & 493.58 \\ \bottomrule
\end{tabular}
\caption{Impact of training data bound identification on the sampling in terms of coverage score}
\label{tab:boxbound_cov}
\end{table}

\subsection {Test case generation}

    \begin{figure*}[]
        \centering
        \subfigure[Coverage-guided Explorer]{\includegraphics[width=0.29\textwidth]{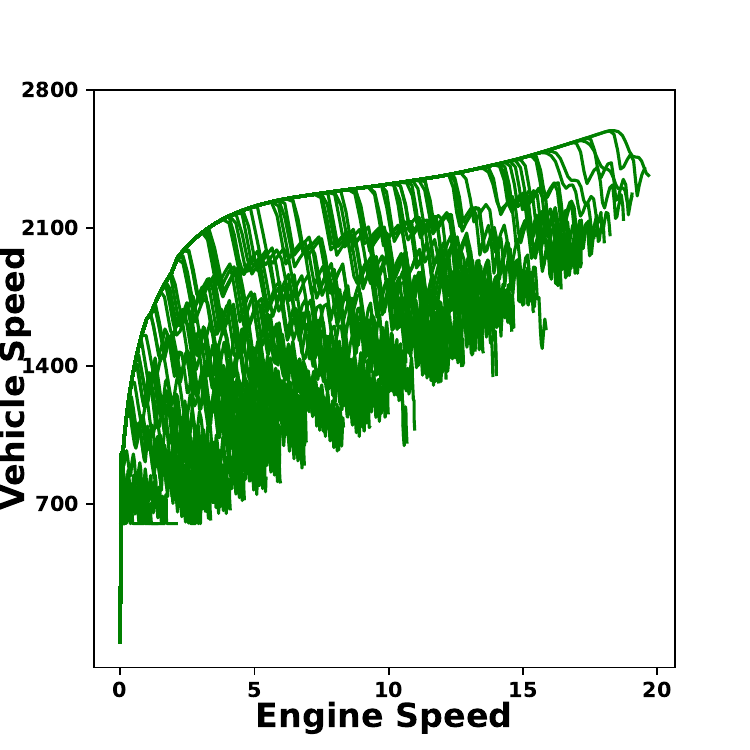}}%\hfill
        \subfigure[Random]{\includegraphics[width=0.29\textwidth]{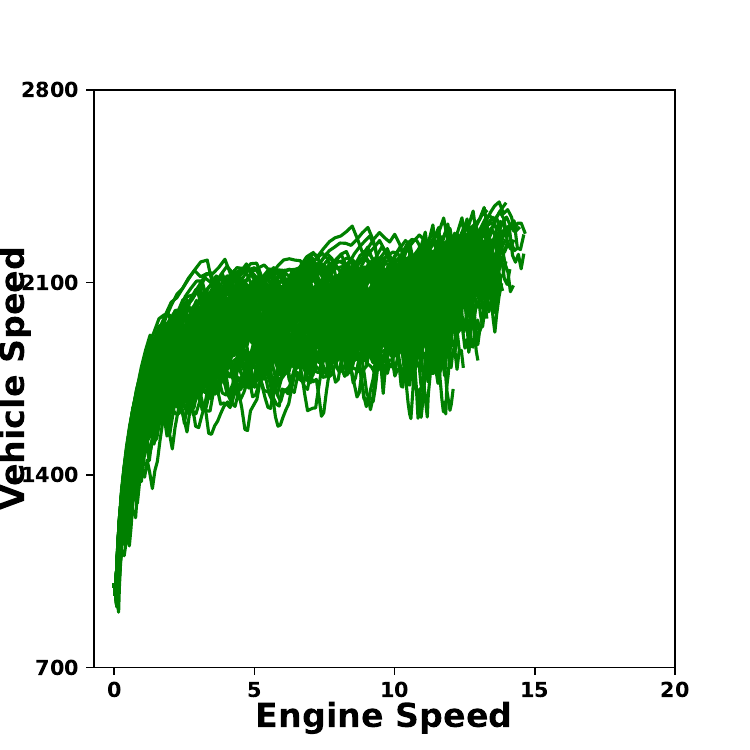}}%\hfill
        \subfigure[S-TaLiRo]{\includegraphics[width=0.29\textwidth]{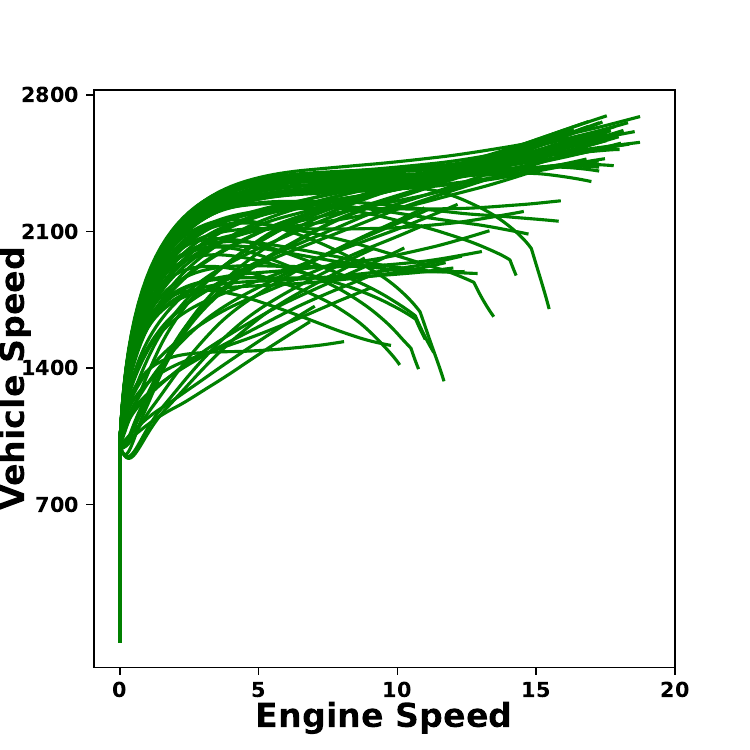}}

        \caption{The state trajectories are obtained by simulating the Automatic Transmission benchmark with evaluation test cases. In this benchmark, the output states include engine speed and vehicle speed.}
        \label{fig:coverage_traj}
    \end{figure*}

The second phase of our work is test case generation. To assess the quality of the generated test cases, we run simulations using them and calculate the CPS coverage score based on the resulting state trajectories.

We performed extensive experiments for test case generation, employing our coverage-guided state space exploration method, a random baseline method, and the S-TaLiRo falsification tool. S-TaLiRo works seamlessly with benchmarks in Simulink and Matlab. Additionally, an adaptable Python toolbox, $\Psi$-TaLiRo, is used for search based test generation in cyber-physical systems developed for Python. In this work we use $\Psi$-TaLiRo for kinematic car and ACAS XU benchmarks and S-TaLiRo for automatic transmission benchmark.

\subsubsection{\textbf{Test Methods}}
We explore different scenarios, attempting to create analogous ones for both the coverage-guided methodology and the falsification tool. We aim to provide similar complexity levels for scenarios in both methods.
%

%\vspace{10}

\subsubsection{\textbf{Objectives}}
In test case generation, we consider different scenarios to determine how to achieve state space coverage. This guides us in defining the cost function for MPC in our coverage-guided methodology or specifying criteria for falsification in S-TaLiRo ($\Psi$-TaLiRo). We explore different scenarios, attempting to create analogous ones for both the coverage-guided methodology and the falsification tool if possible. We aim to provide similar complexity levels for scenarios in both methods:

\begin{itemize}
    \item \textbf{Kinematic Car}: We sample target states using coverage as a guide and minimize the distance between the car position and the target state through the cost function. This approach allows the car to explore diverse state trajectories. 

    In $\Psi$-TaLiRo, we aim for the car, initially positioned at the origin, to move away from origin and explore the state space: $\square(x \neq 0 \land y \neq 0)$

    \item \textbf{ACAS Xu}: Typically, the ownship starts from the origin and the controller issues a clear of conflict control command, causing it to fly straight up without turning unless there's a collision risk. The objective is to test the ownship's neural network controller by placing it in situations where it issues turn advisories. To achieve this, we sample different target states for the intruder aircraft. The cost is defined as the distance between the intruder aircraft's position and the target state. The intruder visits various states, increasing the likelihood of collision or near-collision situations. Consequently, the ownship controller issues turn advisories, prompting the ownship to change directions and explore different states in the state space. 

    In $\Psi$-TaLiRo, we aim for the ownship to depart from the origin to assess its extent of exploration in the state space: $\square( x\_{own} \neq 0)$

    \item \textbf{Automatic transmission}: The target state comprises different combinations of engine speed and vehicle speed. The cost function is designed to guide the system from the initial state to this target state.

    In $\Psi$-TaLiRo, when transitioning to gear 4, there must be no change to any other gear within a 2.5-second interval.
    
    $\square_{[0, 30]} (!gear4 \land X(gear4)) -> X \square_{[0, 2.5]} (gear4))$
\end{itemize}

%\vspace{10}
    
\subsubsection{\textbf{CPS coverage score evaluation}}

We assess state space coverage using the CPS coverage score explained in section \ref{subsec:CPS_Score}. Although another coverage evaluation metric is discussed in related work, we opted for the CPS coverage score. This choice was made because it effectively distinguishes similar states, represents genuine coverage distribution, and involves lower computation overhead.

We tested all methods on the benchmarks mentioned in subsection \ref{subsec:Benchmarks}. The number of test cases generated and the execution time per each test case were consistent across methods for each benchmark, though they varied between benchmarks.  Table \ref{tab:coverage_score} shows that our coverage-guided approach outperforms other methods. To ensure reliability, we conducted tests for each method ten times, reducing the impact of randomness. Table \ref{tab:coverage_score} reports CPS coverage scores in the form of mean and standard deviation, and Fig. \ref{fig:coverage_traj} visually confirms the superiority of coverage-guided test case generation. Each plot illustrates the state trajectories of the generated test cases (projected to $2D$ space).

\begin{table}[]
\centering
\renewcommand{\arraystretch}{1.5} % Increase row spacing
\begin{tabular}{lccc}
\toprule
\textbf{Test Method} & \textbf{Kinematic car} & \textbf{ACAS Xu} & \textbf{Auto Trans} \\
\midrule
\textbf{Coverage Explorer} & $2.45 K \pm 119.0$ & $5.5 K \pm 90.0$ & $0.49 K \pm 14.7$ \\
\textbf{Random} & $0.3 K \pm 12.2$ & $2.3 K \pm 45.7$ & $0.19 K \pm 3.7$ \\
\textbf{STaliro} & $10.39 \pm 5.0$ & $63.5 \pm 33.4$ & $0.36 K \pm 11.2$ \\
\bottomrule
\end{tabular}
\caption{CPS Coverage score mean and standard deviation for 10 execution rounds to eliminate effect of randomness.}
\label{tab:coverage_score}
\end{table}

% \subsubsection{\textbf{Coverage improvement}}
% Assessing coverage improvement over time offers a dynamic view of the testing process. It aids in pinpointing gaps, refining testing strategies, tracking development progress, establishing realistic goals, validating code changes, and ultimately enhance the testing methodology's robustness and reliability. Figure O illustrates the coverage improvement achieved by various test case generation methods across different benchmarks.

\section{CONCLUSION}
\label{sec:conclusion}
We considered CPS test case generation using a methodology focused on state space coverage. 
Our approach involves a coverage-driven, data-driven model training technique for complex black-box CPS and a test case generation method based on coverage enhancement. 
In our model training, we use clustering to reduce similar data traces and allocate test capacity to diverse data, effectively representing the system's state space. We introduced a concept called coverage-guided sampling, where we sample target states from areas with low coverage using state coverage measurement and rejection sampling techniques. 
This method guides the search during model training and test case generation phases, improving coverage. Our approach demonstrates its superiority through benchmark evaluations and comparisons with other methods, emphasizing the importance of integrating state space coverage.

\bibliographystyle{splncs04}
\bibliography{bak, Refrences}
% \bibliography{bak}

\end{document}